\documentclass[doublecol]{epl2} 

\usepackage{amssymb}
\usepackage{amsmath}
\usepackage{color}
\usepackage{graphics, graphicx}
\usepackage{bbold}
\usepackage{psfrag}
\usepackage{mathcomp}
\usepackage{subfigure}
\usepackage{dsfont}

\title{Rigorous mean-field dynamics of lattice bosons: Quenches from the Mott insulator}

\author{Michiel Snoek}

\institute{Institute for Theoretical
Physics, University of Amsterdam, 1090 GL Amsterdam, The Netherlands}

\date{\today}


\begin{document}

\abstract{
We provide a rigorous derivation of Gutzwiller mean-field dynamics for lattice bosons, showing that it is exact on fully connected lattices. We apply this formalism to quenches in the interaction parameter from the Mott insulator to the superfluid state. Although within mean-field the Mott insulator is a steady state, we show that a dynamical critical interaction $U_d$ exists, such that for final interaction parameter $U_f>U_d$ the Mott insulator is exponentially unstable towards emerging long-range superfluid order, whereas for $U_f<U_d$ the Mott insulating state is stable. We discuss the implications of this prediction for finite-dimensional systems.}

\maketitle

\section{Introduction}
Because the energy scales in ultracold quantum gases are much smaller than in solid-state systems, time scales are much larger. This provides the unique opportunity to investigate out-of-equilibrium many-body quantum mechanics on experimentally tractable time-scales. 
Moreover, these systems are well isolated from the environment, such that no decoherence destroys the quantum correlations.
Many fascinating examples have already been published, including
collapse and revival dynamics of strongly correlated bosons \cite{Greiner02}, 
relaxation dynamics in one dimension \cite{Konshita06, Trotzky11},
particle transport of fermions \cite{Strohmaier07} and bosons \cite{McKay08},
diffusion dynamics of strongly correlated fermions \cite{Schneider10}, and
many-body Landau-Zener dynamics \cite{Chen11}.
 
The theoretical description of the dynamics of quantum  many-body systems is very challenging, since already systems in thermal equilibrium require significant effort. 
However, for  lattice bosons a simple, non-perturbative mean-field theory is available, consisting of a mean-field approximation in the hopping part of the Hamiltonian \cite{Fisher89, Rokhsar91,  Sheshadri93, Jaksch98, Oosten01}. Henceforth we will refer to this as Gutzwiller mean-field theory. This approximation can be shown to be exact on fully connected lattices \cite{Fisher89} and in the limit of infinite dimensions \cite{Byczuk08, Hubener09} regardless of the strength of the interaction, which means that the method is fully non-perturbative.
Gutzwiller mean-field theory offers a straightforward extension towards out-of-equilibrium situations \cite{Jaksch02}, 
which has already been applied to many problems:  
creation of a molecular condensate \cite{Jaksch02},
dynamics of the superfluid-insulator phase transition \cite{Zakrzewski05},  
transport in inhomogeneous systems \cite{Snoek07}, 
collapse and revival oscillations \cite{Wolf10},
atom lasers \cite{Hen10},
ramp-up dynamics of the optical lattice \cite{Wernsdorfer10, Natu10},
Bragg scattering \cite{Bissbort10}, 
and trap dynamics \cite{Buchhold10}.
So far this method has been justified by semi-classical arguments, but no rigorous derivation is available, making it questionable in which circumstances this approximation can be trusted.  

In this Letter, we show that the dynamical Gutzwiller equations can be rigorously derived on fully connected lattices, by using the insights recently obtained in Ref. \cite{Sciolla10}.  
This is an intriguing result, because the static Gutzwiller equations are also exact on the fully connected lattice. Moreover,  static Gutzwiller mean-field theory has proven to be a good approximation in three spatial dimensions. Therefore, this results gives confidence that the dynamic Gutzwiller equations are also a good approximation in three spatial dimension.
 
As an application we consider quenches from the superfluid phase towards the Mott insulator phase. Within Gutzwiller mean-field theory the Mott insulating state is trapped: the state is inert when the interaction is decreased towards the superfluid regime. 
Here we investigate the stability of this steady state after the quench, by applying a small perturbation to the Mott state. 
We find that apart from the equilibrium critical interaction $U_c=(3+\sqrt{8})J$, which separates the superfluid (for $U<U_c$) from the Mott insulator (for $U>U_c$), a second dynamical critical interaction $U_d=(3-\sqrt{8})J$ exists, separating  
the regime where the Mott state is stable (for final interaction $U_f<U_d$)
from the regime where the Mott state is unstable with respect to emerging superfluid order (for $U_d<U_f<U_c$). We can rigorously prove this when we use a cut-off on the particle number per site $N_{\rm max} \leq 2$ by using the mapping to the classical system developed in \cite{Sciolla10}. This boundary remains fully intact when removing this constraint, which is shown by explicit time-evolution of the Gutzwiller mean-field equations.
We close by discussing the experimental implications of our findings.

\section{Model}
We consider the homogeneous Bose-Hubbard model, described by the Hamiltonian:
\begin{equation}
\mathcal{H} = - \sum_{i\neq j} J_{ij} \hat b_i^\dagger \hat b_j - \mu \sum_i \hat n_i + \frac{U}{2} \sum \hat n_i (\hat n_i - 1)
\label{ham}
\end{equation}
Here $\hat b_i^{(\dagger)}$ are the annihilation (creation) operators at site $i$ and $\hat n_i = \hat b_i^{\dagger} \hat b_i$  are the corresponding number operators. The particle density is controlled by the chemical potential $\mu$ and the on-site interaction is described by the interaction parameter $U$. $J_{ij}$ contains the hopping amplitudes between the sites $i$ and $j$. 
When applied to a finite-dimensional lattice, we take  $J_{ij} = J/z$ when $i$ and $j$ are nearest neighbors ($z$ denotes the number of neighbors) and zero otherwise. We also consider a fully connected lattice, on which the particles can tunnel from every site to every other site with equal tunnel amplitude $J_{ij} = J/V$, where $V$ is the number of lattice sites. 

\section{Method}
When treated in Gutzwiller mean-field approximation, 
the Hamiltonian in Eq. (\ref{ham}) is replaced by a sum of single-site Hamiltonians $\mathcal{H}_{\rm MF} = \sum_i \mathcal{H}_{\rm MF}^i$ with
\begin{equation}
\mathcal{H}_{\rm MF}^i = - J  \left( \hat b_i^\dagger  \phi + \phi^*\hat b_i  - | \phi|^2 \right) - \mu \hat n_i + \frac{U}{2}\hat n_i (\hat n_i - 1),
\label{ham2}
\end{equation} 
together with the self-consistency relation $\phi = \langle \hat b_i \rangle$. Note that due to the choice of the $J_{ij}$ the form of the mean-field Hamiltonian in Eq. (\ref{ham2}) is independent of the lattice type.
Since the mean-field Hamiltonian is a sum of commuting single-site Hamiltonians, its eigenstates are product states over the lattice sites: $|\Phi\rangle_{\rm MF} = \prod_i |\phi\rangle_i$.  The local state $|\phi\rangle_i$ can be decomposed into Fock states $|n\rangle_i = \hat b_i^\dagger/\sqrt{n!}|0\rangle$: $|\phi\rangle_i = \sum_{n=0}^{N_{\rm max}} c_{n} |n\rangle_i$, where we have assumed a homogeneous state, in which the coefficients $c_n$ do not depend on the site index $i$. We also have explicitly included the cut-off in the particle number $N_{\rm max}$ in the summation.

The \emph{dynamical} Gutzwiller equations follow from the Schr\"odinger equation with the mean-field Hamiltonian (\ref{ham2}) applied to the product state  $|\Phi\rangle_{\rm MF}$: $i \partial_t  |\Phi(t)\rangle_{\rm MF} = \mathcal{H}_{MF}  |\Phi(t) \rangle$, together with the self-consistency equation $\phi(t) = \langle \Phi(t) | \hat b_i |   \Phi(t) \rangle$. Here and in the following we set $\hbar=1$. This leads to time-dependent coefficients $c_n(t)$, which evolve according to the non-linear differential equation:

\begin{eqnarray}
i  \dot c_n(t) &=& -J \Big\{ \phi(t) \sqrt{n} c_{n-1} (t) + \phi(t)^* \sqrt{n+1} c_{n+1} (t) \Big\} \nonumber\\
&& +\Big\{\frac{U}{2} n (n-1)- \mu n \Big\} c_n (t), 
\label{eom1}
\end{eqnarray}
together with 
$\phi(t) =\sum_{n=1}^{N_{\rm max}} \sqrt{j} \; c_{n-1}^*(t)c_n(t)$
These equations can straightforwardly be implemented in matrix-form with low numerical effort, even for large values of $N_{\rm max}$. We exploit this in the second part of the paper, where we investigate quench dynamics.

To make contact with the exact dynamics on the fully connected lattice, we write $c_n = a_n e^{ i \alpha_n}$, where $a_n$ is the absolute value of $c_n$ and $\alpha_n$ its phase. We also introduce $m_n = a_n^2$, which is the probability of having exactly $n$ particles per site, or equivalently, the fraction of sites with exactly $n$ particles.  
From Eq. (\ref{eom1}) we can directly derive the equations of motion for $\alpha_j$ and $m_j$ ($\Delta \alpha_j = \alpha_j - \alpha_{j-1})$:
\begin{eqnarray}
 \dot \alpha_j \hspace{-2mm} & = & \hspace{-2mm} - \frac{1}{2} j (j-1) U + \mu j 
 		\nonumber \\ &&
 + 
J  \sum_k 
	\sqrt{\frac{m_k m_{k-1}}{m_j} k} 
		\Big\{
			\sqrt{m_{j-1} j} \cos(\Delta \alpha_j -  \Delta \alpha_k)
			\nonumber \\ &&
			\quad\quad\quad+\sqrt{m_{j+1}  (j+1)} \cos(\Delta \alpha_{j+1} - \Delta \alpha_k) 
		\Big\},
\label{alphadot}
\\
 \dot m_j\hspace{-2mm}& = &\hspace{-2mm} 
	2 J \sum_k \sqrt{m_j m_k m_{k-1} k} \Big\{
  		\sqrt{m_{j-1} j} \sin(\Delta \alpha_j  - \Delta \alpha_k) \nonumber \\ && 
		\quad\quad\quad- \sqrt{m_{j+1}  (j+1)} \sin(\Delta \alpha_{j+1} - \Delta \alpha_k) 
\Big\}. 
\label{mdot}
\end{eqnarray}

On fully connected lattices the dynamics can be derived without any approximation. Here we closely follow the derivation of  \cite{Sciolla10}: on fully connected lattices ground states must be completely symmetric when sites are interchanged. We can therefore build up a complete set of states contributing to the ground state by considering states with fixed numbers $\{M_i\}$ of sites with $i$ particles, which are fully symmetrized over all sites. 
Later on we will use $m_i = M_i/V$ as the fraction of sites with $i$ particles. Note that this definition of $m_i$  coincides with the one used before in the context of Gutzwiller dynamics.
The states can thus be denoted by $|M_0, \ldots, M_V\rangle$  and the ground state has the form 
$|{\rm gs} \rangle =  \sum_{\{m_i\}} A_{M_0,M_1,\ldots,M_V} |M_0,  \ldots, M_V\rangle$.
Normalization imposes the contraint $\sum_{\{M_i\}} |A_{M_0,\ldots, M_V}|^2 =1$, whereas particle number conservation yields the second constraint 
$\sum_{\{M_i\}} (\sum_i M_i) | A_{M_0\cdots M_V}|^2 =N$, where $N$ is the number of particles. Like in the case of the Gutzwiller mean-field equations, we can put a cut-off on the maximal number of particles per lattice site $N_{\rm max}$. The simplest non-trivial
case is $N_{\rm max} = 2$. Due to the two constraints, the states can then be labeled by the single integer $M_2$ \cite{Sciolla10}. We will make use of this simplification later. 
The symmetry constraints used to determine the ground state also apply when studying time-evolution induced by a quench in the parameter $U$, since this preserves the permutation symmetry of the Hamiltonian.
For general $N_{\rm max}$, the Bose-Hubbard Hamiltonian couples states $|\ldots, M_{j-1}, M_{j},\ldots, M_{k-1}, M_{k}, \ldots \rangle$ to $| \ldots, M_{j-1}+1, M_{j} -1,\ldots, M_{k-1} -1 , M_{k}+1, \ldots \rangle$ with amplitude $W_{j k}$.  Note that $W_{jj}$ are the diagonal elements and $W_{j,j+1}$ couples   $|\ldots, m_{j-1}, m_{j}, m_{j+1}, \ldots \rangle$ to $| \ldots, m_{j-1}+1, m_{j} -2,\ldots, m_{j+1}+1, \ldots \rangle$.
The Schr\"odinger equation for the amplitudes $A_{M_0\cdots M_V}$ , expressed in terms of the $W_{jk}$ therefore yields:
\begin{eqnarray}
&& i \dot A_{M_0\cdots M_{j-1} M_j \cdots M_{k-1} M_k \cdots M_V} = \\ 
&& \quad \sum_{j k}  W_{j k} A_{M_0\cdots (M_{j-1}+1) (M_j -1) \cdots   (M_{k-1} -1) (M_k+1) \cdots M_V} \nonumber
\end{eqnarray}

We now consider the thermodynamic limit $V\rightarrow \infty$. In this limit the transition amplitudes $W_{kl}$ assume a simple form: 
\begin{eqnarray}
W_{j k} &=& \left( \sum_i \left\lbrack \frac{U}{2}  M_j j (j-1) - \mu \; M_j j \right\rbrack \right) \delta_{jk} - 
\nonumber \\ &&  J \sqrt{M_j M_{j-1} M_k M_{k-1}} \sqrt{jk},
\end{eqnarray}
such that $W_{jk} = W_{kj}$. Turning now to the $\{m_j\}$ variables, which become continuous in this limit, the Schr\"odinger equation turns into:
$$
i \dot A(\{ m_j \}) =  \sum_{jk}  W_{jk} \cos(p_j -p_{j-1} - p_{k-1}+p_k) A(\{ m_j \}) \nonumber
$$
where we have introduced for each $m_j$ an associated momentum $p_j = -i \partial_{m_j}/V$ in exactly the same manner as was derived in  \cite{Sciolla10} for $N_{\rm max} = 2$. 
In the thermodynamic limit $m_j$ and $p_j$ commute and
we obtain classical equations of motion $m_j = \partial H/\partial p_j$ and $p_j = -\partial H/\partial m_j$ for the classical Hamiltonian
\begin{eqnarray}
H &=& \frac{U}{2} \sum_j m_j  j (j-1) - \mu \sum_j m_j j -
\\ && \nonumber
J \sum_{j,k} \sqrt{m_j m_{j-1} m_k m_{k-1}} \sqrt{jk} \cos(\Delta p_j - \Delta p_k)
\end{eqnarray}
It is now straightforward to see that these classical equations of motion exactly coincide with the Gutzwiller equations of motion in Eqs. (\ref{alphadot}, \ref{mdot}) when we identify the phase $\alpha_j$ with the momentum $p_j$.

This constitutes the proof that Gutzwiller dynamics is exact on fully connected lattices. Therefore it is 
 a controlled mean-field method with the same range of validity as the static Gutzwiller equations which are also exact on fully connected lattices \cite{Fisher89}.  
 
The fact that the static and dynamic Gutzwiller equations are both exact on fully connected lattices, i.e. in infinite dimensions, gives support for the idea that the accuracy of this mean-field method in finite dimensional lattice is comparable for both cases. The accuracy of the static mean-field theory has been extensively tested, because the results can be compared with numerically exact results from Quantum Monte Carlo simulations \cite{QMC3d, QMC2d}.  One can for instance compare the prediction for the critical interaction $(U/J)_c$ for the superfluid-Mott insulating transition. This yields reasonably good quantitative agreement in three dimensions. 
In two dimensions the agreement is merely qualitative. 
Also for inhomogeneous systems Gutzwiller mean-field predictions have been compared with Quantum Monte Carlo data, yielding a similar agreement as for the homogeneous case \cite{Hen10b}.
This progressively good accuracy of the mean-field theory when the dimension is increased, can be understood by viewing the static Gutzwiller mean-field equations as the leading terms in a $1/z$ expansion \cite{Hubener08, Snoek10}. Including the second order terms yields Bosonic Dynamical Mean-Field Theory \cite{Byczuk07, Hubener08, Anders10, Snoek10, Anders11}, for which almost exact agreement with the Quantum Monte Carlo results has been obtained \cite{Anders10, Anders11}. 
Several efforts have already been made to go beyond the mean-field theory for the out-of-equilibrium situation as well \cite{Trefzger11, Navez10}

It is worth noting that although the mean-field theory turns out to be a good approximation in sufficiently high dimensions, the ground state description does not include any entanglement between different lattice site: in mean-field the states are perfect product states.
 
The derivation  presented here is clearly independent of the strength of the interaction parameter and hence valid both in weak and strong coupling.
The equations are moreover valid for all types of dynamics: since no adiabatic assumptions have been made, the Gutzwiller dynamics can be applied for slow, quasi-adiabatic dynamics, as well as for fast dynamics. An example of the latter is the quench dynamics we turn to now.

\begin{figure}
\includegraphics[width=8cm, height=7cm]{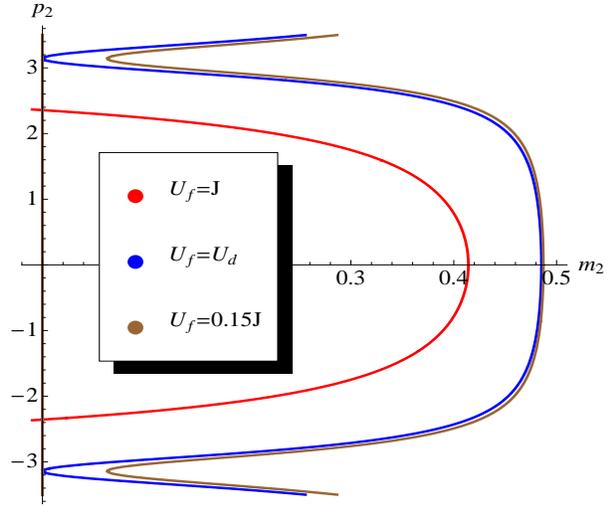}
\caption{Classical paths for $N_{\rm max} = 2$ with $H=0$ for various values of $U_f$.}
\label{classicalpaths}
\end{figure}

\section{Quenches from the Mott insulator}
We study the situation in which after preparing the ground state for some value of $U=U_i$, the interaction is suddenly changed to $U=U_f$ at time $t=0$ and the subsequent time-evolution of the double occupancy $m_2(t)$ and the condensate fraction $N_c(t) = \sum_{i\neq j} \langle b_i^\dagger b_j \rangle/V^2 = |\phi(t)|^2$ is studied.  
We study the case that the particle density $n=N/L = 1$.
As is well known, the equilibrium properties are then that within Gutzwiller the system is superfluid for $U<U_c = (3+\sqrt{8}) J$ and Mott insulating for $U>U_c$. This last state is particularly featureless withing Gutzwiller: it is a perfect product state with exactly one particle per site.

In previous work \cite{Sciolla10} quenches within the superfluid phase were studied. A dynamical transition was found between small and large quantum quenches.

Here we study quenches from the Mott insulating state, i.e. we choose $U_i > U_c$ \cite{Trefzger11, Altman02, Polkovnikov02}. 
This might seem problematic, because the Mott state is completely inert within mean-field theory: when the system is in the Mott insulating state, the system will remain in this state for all $U_f$. We overcome this problem here, by analyzing the stability of the Mott state after a quench: a small perturbation is added to the perfect Mott state, which can drive the system away from the insulator and thus prove the insulator unstable. We note that within mean-field this perturbation might seem artificial; however, quantum fluctuations beyond mean-field, which are always present in finite-dimensional systems, naturally provide this perturbation.

We now first analyze the situation for $N_{\rm max} = 2$. In this case, the equations of motion become very simple, because of the conservation of particle number and norm. Using this, one obtains coupled classical equations of motion for $(m_2, p_2)$ with $H = U m_2 - J( 1-2 m_2)m_2 (3+\sqrt{8}\cos(p_2))$ \cite{Sciolla10}. Moreover $N_c(t) = m_2(t)(1-2m_2(t))(3 + \sqrt{8} \cos p_2(t))$, and hence it is only nonzero if $m_2$ is nonzero and proportional to $m_2(t)$ for small $m_2(t)$. Therefore we focus on the time evolution of $m_2(t)$ in the following. 

For the Mott insulating state $m_2 = 0$ and hence the energy $H=0$. Since energy is conserved, the system has to evolve along paths with $H=0$. These are shown in Fig. \ref{classicalpaths} for various values of $U_f$. The path with $m_2 = 0$ is always a solution, and when initially $m_2(0)=0$, this will remain true for all later times. (This is true in the thermodynamic limit; for finite systems on the fully connected lattice the system can escape from this state on a timescale diverging with the system size as $\log(V)$  \cite{Sciolla10}.) 

\begin{figure}
\begin{center}
\hspace{-1cm}\includegraphics[scale=.7]{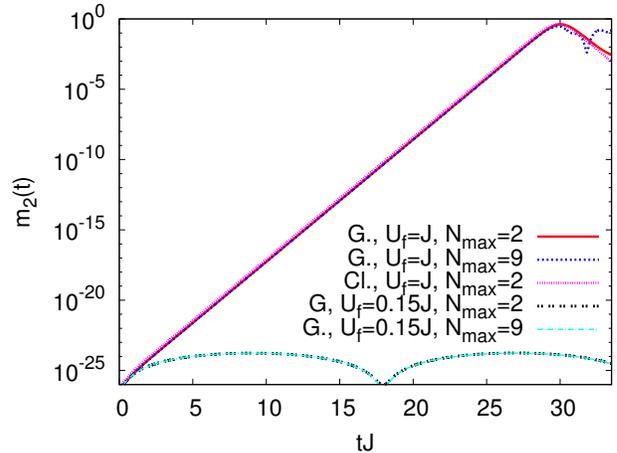}

\vspace{-.5cm}

\end{center}
\caption{Time evolution of the double occupancy $m_2(t)$ starting from the Mott insulator with a tiny perturbation quenched towards $U_f > U_d$ ($U_f = J$) and $U_f<U_d$ ($U_f = 0.15J$). Shown are results from Gutzwiller dynamics (G.) with different $N_{\rm max}$, and classical dynamics (Cl.) for $N_{\rm max} = 2$.  }
\label{timeevolution}
\end{figure}

\begin{figure}
\hspace{-1cm}\includegraphics[scale=.7]{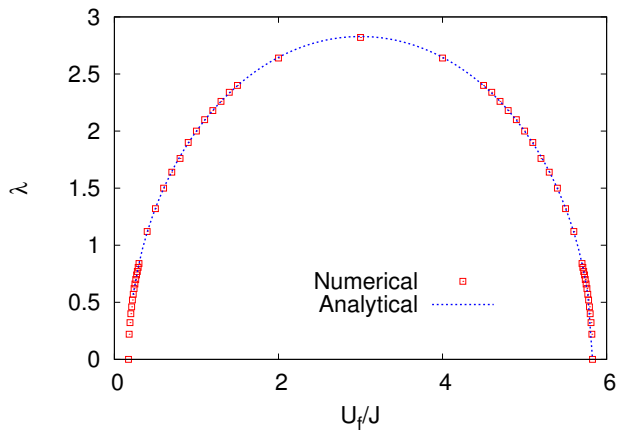}

\vspace{-.5cm}

\caption{Exponent $\lambda$ of the exponential growth of  $m_2(t) \propto e^{ \lambda t J}$ as a function of $U_f$, both obtained by numerical fitting (squares) and the analytic expression (line).}
\label{exp_fit}
\end{figure}

We now want to investigate the stability of this solution: what happens if $m_2(0)$ is slightly positive? 
To answer this question we look at the other branch with $H=0$ which is characterized by $m_2>0$. 
Now a fundamental difference appears between $U_f<U_d$ and $U_f>U_d$ (where $U_d = (3-\sqrt{8})J$): in the former case this branch is fully disconnected from the $m_2 = 0$ curve. 
This implies that paths with initial conditions with $m_2 \ll 1$ remain infinitesimally close to $m_2 = 0$  
and the Mott state is dynamically stable under small perturbations. An important direct consequence of this is that the system does not thermalize: because the system is completely static, there is no evolution towards a state that could be characterized by a fictitious temperature.

In contrast, when $U_f>U_d$, the two branches cross, such that in this case solutions with $m_2(0)>0$ will follow a curve on which $m_2(t)$ grows macroscopically large. Hence the Mott state is unstable for this case. We verify this numerically by time-evolving the classical equations of motion and the Gutzwiller equations for initial states that are slightly perturbed away from the pure Mott insulator. 
This fully confirms this picture: for $U_f>U_d$ the double occupancy grows exponentially, whereas for $U_f<U_d$ it remains small and oscillates. 
Note that these oscillations are fully within the regime of an infinitesimal $m_2$ and hence a stable Mott insulator. They correspond to the time evolution along the path infinitesimally close to the line $m_2=0$,  which still involves dynamics of $p_2$, reflecting itself in oscillations in $m_2$ as well.

This is shown in Fig. 
\ref{timeevolution}, in which also results with higher $N_{\rm max}$ are shown.  
We observe that the initial exponential growth is independent of $N_{\rm max}$: only after the superfluid order parameter has grown macroscopically large, differences due to different values of $N_{\rm max}$ are visible. We can fit the double occupancy with an exponential fit over many orders of magnitude: $m_2(t) \propto e^{ \lambda t J}$.
The exponent $\lambda$ from this fit is shown in Fig. \ref{exp_fit}.
It can be derived analytically by investigating the classical equations of motion for $N_{\rm max} = 2$. The time-evolution starts with $m_2 = p_2 = 0$. First the system evolves along the $m_2 = 0$-line, until $\dot  p_2 = 0$ at $\cos p = (U_f - 3J)\sqrt{8}J.$  Then the curve sharply bends and moves with constant $p_2$ 
driven by the linearized equation of motion $$\dot m_2 = \sqrt{8} J m_2 \sin p_2 = m_2 \sqrt{(U_f - U_d)(U_c - U_f)}.$$  This yields the observed exponential increase, with exponent $\lambda = \sqrt{(U_f - U_d)(U_c - U_f)}$, which exactly fits the numerical fits in Fig. \ref{exp_fit}. We observe a remarkable symmetry around $U_f = 3 J$ in this picture. The exponent vanishes like a square-root when $U_f$ approaches $U_c$ and $U_d$, like one expects from mean-field dynamics.
It is worth noting that the critical point does not shift in this case, as generally expected when studying quench dynamics in dimensions $d \geq 2$ \cite{Gambassi10}. This is because the mean-field Mott insulating state we start from is a perfect product state, in which the correlation length exactly vanishes.

\section{Conclusions and experimental implications}
We have shown that Gutzwiller mean-field theory is exact on fully connected lattices, i.e. in infinite dimensions, irrespective of the interaction strength or the question whether the dynamics is slow or fast. 
We applied this mean-field formalism to quenches in the $U/J$-ratio from the Mott insulator to the superfluid side of the phase diagram. Although the pure Mott insulator is always a steady state, we found that infinitesimal fluctuations make this state exponentially unstable towards emerging long-range superfluid order for $U_f>U_d$.

Whereas this is a rigorous statement on the fully connected lattice, we expect this dynamical transition to be present on lattices with finite connectivity as well. Especially for three-dimensional lattices, where quantum fluctuations only lead to small quantitative corrections to the static Gutzwiller predictions, we expect the Gutzwiller dynamics to be a good approximation and the dynamical transition to be present. 
A second argument for this comes from the consideration of the special point $U_f = 0$.  In this case the Hamiltonian driving the time-evolution consists only of the hopping part, which commutes with operators of the form $b_j^\dagger b_i$ measuring long range phase coherence. Hence it is a rigorous statement that for $U_f = 0$ no superfluid coherence can build up and the Mott insulating state is stable for this point. 
Whereas $U_d$ will be shifted away from the mean-field value in finite dimensional lattices (like also the value of $U_c$ is changed), 
this  argument thus implies that $U_d \geq 0$.
However, even the case $U_d=0$ would be interesting, because in principle it is also possible to quench to attractive interactions. 

Another important aspect that changes in finite dimensions, is that the Mott insulating state is no longer featureless: virtual particle hole-pairs lower the energy proportional to $J^2/U_i$. This naturally provides the perturbation from the perfect product state that is needed to trigger the exponential build-up of superfluid order, but it also means that the exact phase boundary can only be probed for initial $U_i/J \gg 1$. 

The existence of a dynamical transition can experimentally be investigated by using ultracold atoms in optical lattices, in which the ratio $U/J$ can easily be quenched by modifying the depth of the optical lattice or by means of Feshbach resonances. In this latter case it is also possible to quench to attractive interactions.
The occurrence of long-range superfluid order is directly visible in time-of-flight experiments, in which the phase coherence is probed, and also the amount of double occupancies can be directly observed \cite{Jordens08}. 

However, the presence of a trapping potential, leading to an inhomogeneous system, complicates the situation. This is because of two reasons: after a quench in the $U/J$-ratio also particle transport sets in, such that the local energies and local particle numbers are no longer conserved. Secondly, if the Mott insulator is surrounded by a superfluid, phase coherence is build up from the outside. 
The sharp dynamical transition is therefore best visible in a flat potential, as realized in \cite{Schneider10}.

This work was supported by the Nederlandse Organisatie voor Wetenschappelijk Onderzoek (NWO).

\end{document}